\documentclass{aa} 

\usepackage{graphicx}
\usepackage{txfonts}

\usepackage{xcolor}

\definecolor{stefan}{rgb}{0.86, 0.08, 0.24}

\usepackage{siunitx}
\usepackage{textcomp}

\defcitealias{Meingast-II:2019}{Paper~II}

\usepackage[toc,page]{appendix}

\usepackage[colorlinks=true,linkcolor=blue,citecolor=blue]{hyperref}%

\begin{document} 
\subtitle{IV. Meingast 1\thanks{In our original discovery paper we did not name the stream. The authors of the first follow-up paper \citep{Curtis:2019} contacted us regarding a name for the structure but did not agree with our proposed name and decided on their own to name the system the Pisces-Eridanus stream. Their chosen name, however, not only does not capture the true size of stream (the stream stretches across at least 10 constellations and likely extends beyond these) it is ambiguous as it can lead to confusion with the Pisces moving group \citep{2018MNRAS.473.2465B}. In general, given the number of new streams being found by Gaia and the finite number of constellations, it seems appropriate to move away from using constellations to name streams (e.g., \citealp{Ibata2019}). An unambiguous remedy to this particular situation is to name the stream after the original discoverer, which we do in this paper, naming the structure Meingast 1.}: the most massive stellar stream in the solar neighborhood}
\title{Extended stellar systems in the solar neighborhood}
   \author{S. Ratzenb\"ock
          \inst{1}\fnmsep 
          \and
          S. Meingast \inst{2}
          \and
          J. Alves\inst{1,2,3}
          \and
          T. M\"oller\inst{1,4}
          \and
          I. Bomze\inst{1,5}
          }
   \institute{Data Science @ University of Vienna,
                W\"ahringer Straße 29, A-1090 Vienna\\
               \email{sebastian.ratzenboeck@univie.ac.at} 
         \and
             University of Vienna, Department of Astrophysics, T\"urkenschanzstrasse 17, 1180 Wien, Austria
         \and Radcliffe Institute for Advanced Study, Harvard University, 10 Garden Street, Cambridge, MA 02138, USA
         \and University of Vienna, Faculty of Computer Science, W\"ahringer Straße 29/S6, A-1090 Vienna
         \and University of Vienna, ISOR/VCOR, Oskar-Morgenstern-Platz 1, A-1090 Vienna
             }


 
  \abstract
  {Nearby stellar streams carry unique information on the dynamical evolution and disruption of stellar systems in the Galaxy, the mass distribution in the disk, and provide unique targets for planet formation and evolution studies. Recently, Meingast et al. 2019 found a nearby ($\sim$100 pc), $120^\circ$ stellar stream with a length of at least 400 pc.}
  {We revisit the stream discovered in Meingast et al (2019) to search for new members within its currently known 400 pc extent, using Gaia DR2 data and an innovative machine learning approach.}
  {We use a bagging classifier of one-class Support Vector Machines with Gaia DR2 data to perform a 5D search (positions and proper motions) for new stream members. The ensemble is created by randomly sampling 2.4 million hyper-parameters realizations admitting classifiers which fulfill a set of prior assumptions. We use the variable prediction frequency resulting from the multitude of classifiers to estimate a stream membership criterion which we use to select high fidelity sources. We use the HR diagram and the Cartesian velocity distribution as test and validation tools.}
  {We find about 2000 stream members with high-fidelity, or about an order of magnitude more than previously known, unveiling the stream's  population across the entire stellar mass spectrum, from B-stars to M-stars, including white dwarfs. We find that, apart from being slightly more metal-poor, the HRD of the stream is indistinguishable from that of the Pleiades cluster. For the mass range at which we are mostly complete, $\sim$0.2 M$_\odot$ $<$ M $<$ $\sim$4 M$_\odot$, we find a normal IMF, allowing us to estimate the total mass of stream to be about 2000 M$_\odot$, making this relatively young stream by far the most massive known. In addition, we identify several white dwarfs as potential stream members.}
  {The nearby Meingast 1 stream, due to its richness, age, and distance, is a new fundamental laboratory for star and planet formation and evolution studies for the poorly studied gravitationally unbound star-formation mode. We also demonstrate that One-Class Support Vector Machines can be effectively used to unveil the full stellar populations of nearby stellar systems with Gaia data.}

\maketitle


\section{Introduction}
Coherently moving groups of stars in the Milky Way are unique laboratories to coherently study a large variety of astrophysical processes. For instance, the similar birth conditions in nearby moving groups have provided much insight into individual stellar properties \citep[e.g.][and references therein]{Torres2008,Gagne2014,Riedel2017}. Moreover, while older stellar systems experience mass loss due to the gravitational interaction with the Galaxy's gravitational potential \citep[e.g.][]{Hyades_Meingast, Hyades_Roser}, young co-moving groups can give important clues on the governing star formation processes in the Milky Way.

Recently \cite{Meingast-II:2019}, the second installment in this series (hereinafter referred to as \citetalias{Meingast-II:2019}), discovered a 120$^\circ$ stellar stream that is currently traversing the immediate solar neighborhood at a distance of only $\sim$\SI{100}{pc}. In this paper, the authors determined the age of the system to be \SI{1}{Gyr}. Their assumption was mostly based on the presence of a single star in their selection, namely the subgiant 42 Ceti. Shortly after the stream's discovery, \citet{Curtis:2019} determined stellar rotation periods of stream members to be very similar to stars in the Pleiades. Their application of gyrochronolgy thus sets the age of the stream to be close to \SI{120}{Myr}, implying that the star 42 Ceti is likely an unfortunate interloper. 

The search criteria in \citetalias{Meingast-II:2019} was based on the 3D space velocities in a cylindrical coordinate frame derived from astrometric measurements provided with the second Gaia data release (Gaia DR2; ~\citealt{Gaia:2016}, \citeyear{Gaia:2018b}). While space velocities provide a robust estimate on membership, evaluating 3D motions of stars requires radial velocity measurements. This requirement limits substantially the identification of members to a small subset of Gaia DR2, specifically to stars with $G\lesssim$\SI{13}{mag}, which in the case of Meingast~1 translate to stellar masses between $\sim$0.5 and 1.5 M$_\odot$. 


The goal of this paper is to unveil the stellar population of the Meingast~1 stream, from B stars down to mid M stars, or the completeness limit of the Gaia DR2 data. To this end, we apply state-of-the-art machine learning tools, where we use the previously identified members as a training set. The structure of this paper is as follows: In Sec.~\ref{sec:data} we present the data used for the analysis. Sec.~\ref{sec:method} summarizes the method used to select potential stream member sources from the Gaia DR2 data set. Finally, in Sec.~\ref{sec:results} we present a final high fidelity source catalogue\footnote{The full source catalogue described in Table~\ref{table:source_list} is only available at the CDS via anonymous ftp.} on which we determine the age and mass of the Meingast 1 stream.


\section{Data}
\label{sec:data}
For the analysis we used the five-dimensional (5D) position ($\alpha$, $\delta$, $\varpi$) and velocity ($\mu_{\alpha}$, $\mu_{\delta}$) information, provided by Gaia DR2. Following the data selection in~\citetalias{Meingast-II:2019} we preferred distance estimates provided by~\cite{Bailer-Jones:2018}. The distance limit of the stellar sample is kept at $\le 300$ pc in accordance to~\citetalias{Meingast-II:2019}. This is motivated by the choice of our classifier which predicts member stars within the limits of the previously determined extent of the stream. Furthermore, the subsequently described method works independently from quality criteria. Therefore, quality filters are only applied for visualisation purposes. This selection results in data set of $18,692,951$ total stars.

In~\citetalias{Meingast-II:2019} the sources are extracted in a six dimensional parameter space spanned by three spatial ($X$, $Y$, $Z$) and three velocity dimensions ($v_r, v_{\phi}, v_z$). Specifically, the velocities were represented in a galactocentric cylindrical coordinate system to better represent the bulk motion stars. Consequently, the source identification in~\citetalias{Meingast-II:2019} depended on radial velocity measurements which are scarce in Gaia DR2. Within the search region of $300$~pc about $95\%$ of all sources in the catalog are, therefore, not taken into account in~\citetalias{Meingast-II:2019} due to missing radial velocity data.


\section{Member selection}
\label{sec:method}
As mentioned above, the bulk of Gaia DR2 catalog sources were not used in the original member identification of the stream in ~\citetalias{Meingast-II:2019}. Omitting the radial velocity component yields a much more complete source list but at the same time limits any analysis to projected tangential velocities given by the proper motion measurements. While members of spatially confined star clusters can be identified reliably in proper motion space, the recently discovered stream encompasses at least \SI{120}{\degree} on sky. This large extent introduces significant projection effects in tangential velocities, posing a non-trivial problem for member identification in five dimensions.


\subsection{Supervised member selection}


To avoid the difficult task of clustering in the 5D position and proper motion space we pursue a supervised approach based on One-Class Support Vector Machines (OCSVM;~\citealt{Scholkopf:2001}). Instead of finding a decision boundary between distinct groups in the training sample like a typical SVM~\citep{Cortes:1995}, an OCSVM constructs a decision surface that attains a maximum separation between the training samples and the origin. Consequently, the algorithm infers the properties of the input samples by enclosing the support of its joint distribution with a hyper-surface during the training process. Depending on the position of unseen data points\footnote{Stars in the data set are represented as points in a 5D space with three position axes and two proper motion axes constituting the so called feature space. Thus, in a machine learning context we refer to stars in the data set as points in a feature space.} to this surface, a trained predictor acts as a binary function which groups new example points as either resembling the previously seen training data or not. We aim to estimate the extent of the stellar stream by using the OCSVM algorithm and the already classified sources from~\citetalias{Meingast-II:2019} as a training set. Subsequently, we predict the membership of unseen stars to the stream within a $300$~pc sphere around the sun (see~Sec.~\ref{sec:data}). In order to find a model which is capable of providing a physically meaningful characterisation of the stellar stream in the 5D feature space, the corresponding hyper-parameters of the OCSVM classifier have to be set sensibly.
\subsection{Parameter tuning}
\label{sec:hp_tuning}
We make use of the \textit{libsvm}~\citep{LIBSVM:2011} OCSVM implementation which features two main hyper-parameters for the RBF-kernel\footnote{We conclude from extensive hyper-parameter searches that the RBF kernel always outperformed the alternative options. Hence we omit the description of other kernel types in this section.}, $\gamma$ and $\nu$. The parameter $\gamma$ defines a region of influence of the support vectors selected by the model. The variable $\nu$ controls the fraction of possible outliers as well as the fraction of support vectors. Thus, $\gamma$ and $\nu$ are crucial hyper-parameters which define the shape of the enveloping hull. 

Additionally, these parameters and subsequently the classifier shape depend on the input variable range. Since the parameter $\gamma$ describes a support vectors region of influence different feature ranges lead to a varying model flexibility within each input variable. To mitigate an asymmetric feature weighting, a common approach is to standardize each input variable to a common variance by dividing each feature by its standard deviation. However, as we are dealing with a combined feature space of position and proper motion information a certain weighting towards one of the two feature spaces might be beneficial to properly characterize the joint probability of stream members. Consequently, after scaling the features to unit variance we add an additional hyper-parameter $c_x/c_v$. This parameter describes the scaling fraction between positional and proper motion features. When $c_x/c_v=1$ the variance in both feature spaces is the same. In practice we set $c_v=1$ and vary $c_x$ within a certain range.

As we choose a classifier via a set of hyper-parameters we have to be aware of existing contamination in the training set (estimated to amount a few percent in~\citetalias{Meingast-II:2019}). Additional selection biases caused by the original clustering and parameter choice which influence the final obtained stream selection should be considered. Therefore, only crude estimates about the true joint distribution of the sources in 5D are possible. Nevertheless, we have information about the resulting classifier shape which limits the space of possible solutions. First, based on the number of missing radial velocity measurements we estimate that the total number of member stars should roughly increase twenty-fold. Second, due to a lack of a better description we estimate that the true extent is comparable to the original selection in~\citetalias{Meingast-II:2019} which found that the stream is roughly prolate spheroidal with a length of about $400$~pc and an equatorial diameter of about $50$~pc.

A trained classifier has to be able to capture these prior assumptions. Therefore, we use the above mentioned characteristics to eliminate predictions which seem unfit to describe the stellar stream in 5D. Since we cannot infer the true joint distribution from the available stream members and our prior assumptions entail some allowable margin of variation, the model parameters cannot be tuned to optimal values. Instead, we aggregate the predictions of multiple models which conform to our prior assumptions into an ensemble of OCSVMs. This procedure is referred to as bootstrap aggregating, also known as \textit{bagging}~\citep{Breiman:1996}. A benefit of using multiple aggregated classifiers, in comparison to one single model, is an improvement in prediction stability. Due to its variance reducing ability, bagging has been successfully applied especially to noise-prone classifiers, whose predictions vary significantly with small variations in the training data. In~\cite{Grandvalet:2004} the author suggests that bagging systematically reduces the influence of outlier samples in the training data. Furthermore, by bundling together multiple models, a notion of stability for each star is obtained as different regions of the 5D training space will have varying prediction frequencies. Ideally, the ensemble of classifiers has a higher prediction frequency towards the center region of the stellar stream (in 5D) where sources are less likely to be randomly selected field stars. Bagging, therefore, automatically creates a hierarchy from more robust to less robust stream members which reduces prediction variance compared to a single classifier.

\begin{figure}[t]
\centering\includegraphics[width=0.8\linewidth]{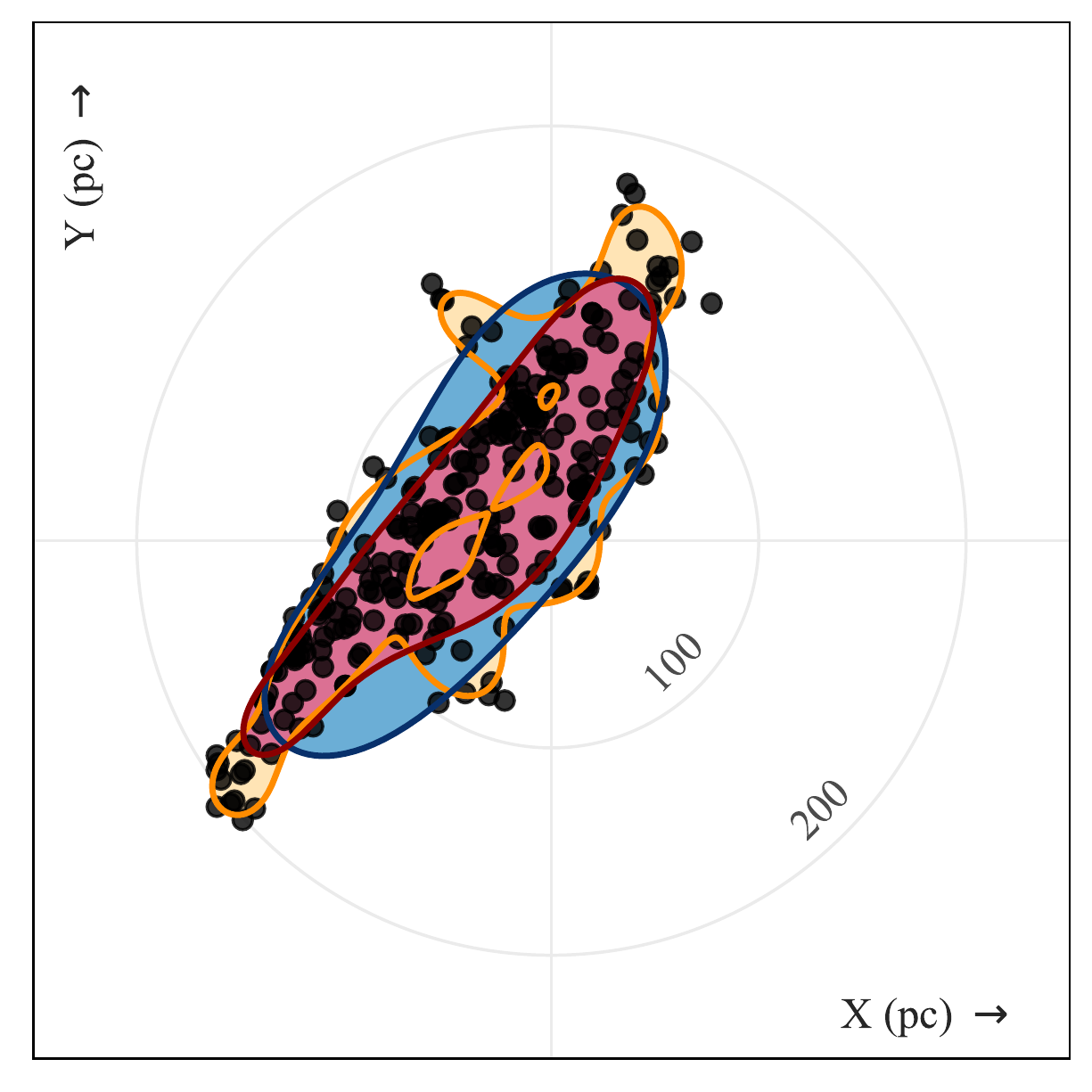}
\caption{Schematic figure illustrating the effect of different hyper-parameters on the classifier shape in the galactic X-Y plane. Black points represent the training set whereas the colored shapes depict the bounding surfaces of individual OCSVM classifiers trained with a different set of hyper-parameters. The unification of multiple classifiers results in an ensemble classifier where overlapping bounding regions result in different levels of stability.}
\label{fig:ensample_schema}
\end{figure}

A schematic illustration of a small ensemble classifier is shown in Fig.~\ref{fig:ensample_schema}. The black scatter points represent the training set whereas the colored shapes depict the bounding surfaces of individual OCSVM classifiers trained with different sets of hyper-parameters. The unification of multiple classifiers results in an ensemble classifier where overlapping bounding regions result in different levels of prediction frequency.

The final bagging predictor is obtained in a two step process: First, the actual training phase and, second, the validation phase, which rejects models that do not represent our expectations well. In the learning phase (see App.~\ref{app:training} for more details)  the model is trained using 10 fold cross validation on a random set of hyper-parameters ($\gamma_{i}$, $\nu_{i}$, $(c_x/c_v)_i$). Before deploying the classifier on the full data set we filter out models below a mean accuracy score of $0.5$, or a standard deviation above $0.15$ across the hold out sets. Models passing this filter criterion enter the validation phase which assess the classifiers capability to capture our prior assumptions about the distribution and quantity of predicted sources. We require the model to comply with the following criteria:

\begin{enumerate}
\item The number of predicted stream members $N_s$ must not exceed a physically sensible range which is limited to $N_s \in [500, 5000]$.
\item The extent of the predicted stream members in position and proper motion space must be similar to the original ones.
\item The cylindrical velocity distribution of the stream members must not deviate too much from the training sample distribution.
\end{enumerate}
For a full description on the implementation of these three validation criteria see App.~\ref{app:validation}.

Eventually, these criteria yield model solutions which capture our prior assumptions about the distribution of the predicted sources. The final model ensemble is subsequently constructed by iterating through $2.4$ million random realizations of ($\gamma_{i}$, $\nu_{i}$, $(c_x/c_v)_i$) within their respective range and storing the individual predictors which fulfill the validation criteria. Altogether, the final classifier ensemble consists of a total of $8515$ classifiers which have passed the validation steps. Fig.~\ref{fig:hp_selected} shows the distribution of accepted models with respect to the hyper-parameters $\nu$, $\gamma$, and $c_x/c_v$.

\begin{figure*}[ht]
\centering\includegraphics[width=0.95\textwidth]{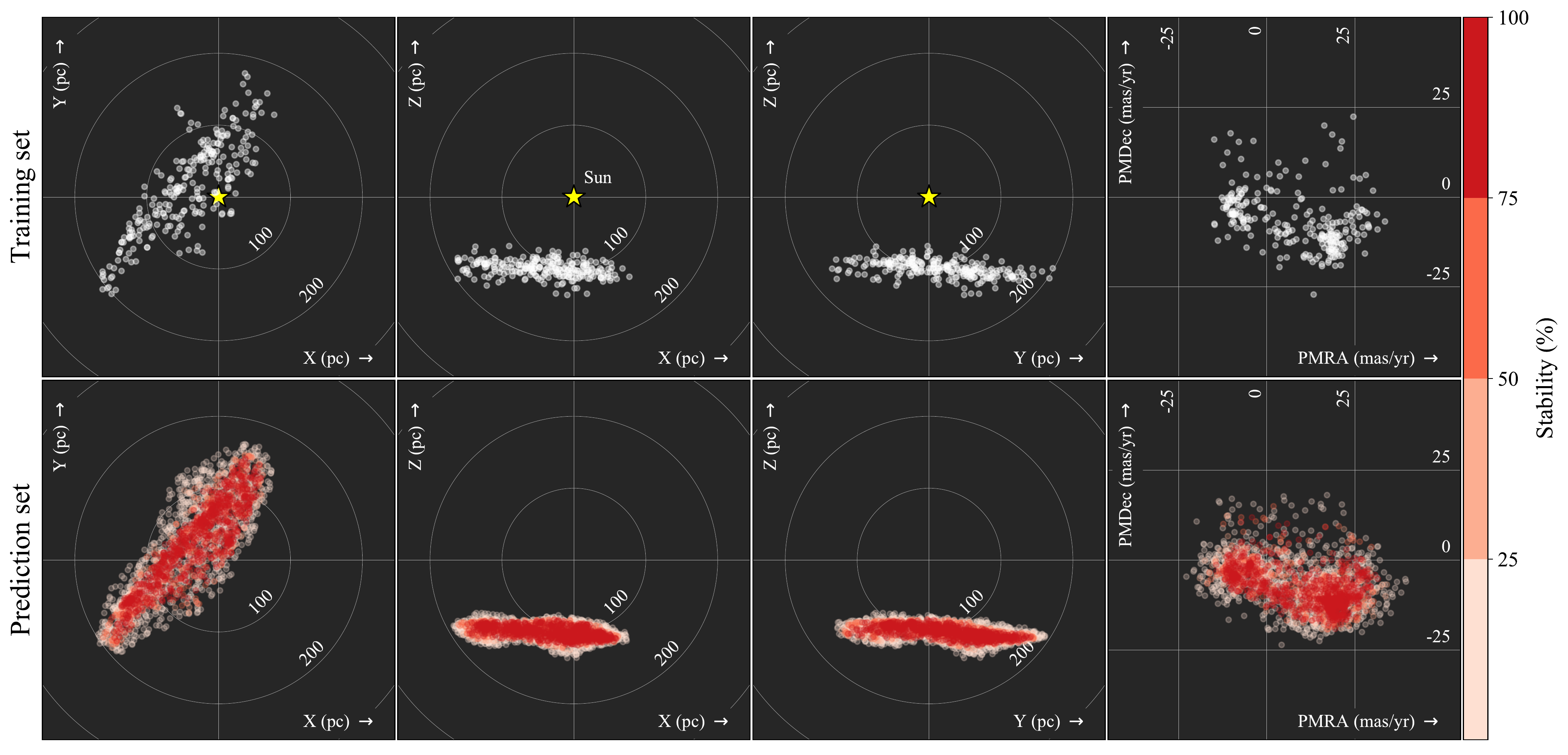}
\caption{Positional and proper motion projections of the training and prediction set are displayed in the first and second row, respectively. Using a quality pre-selection (see App.~\ref{app:cut}) we find a total of $2567$ member stars (bottom row), compared to $256$ in the training set (top row). The color information highlights the stability of a given star which tends to grow towards the central regions of the stream.}
\label{fig:xyz_pm_scattermatrix}
\end{figure*}

\subsection{Limitations and caveats}
Any supervised model based on OCSVMs is limited by the provided training data because the shape of the decision surface is determined by the input training set.

As suggested in~\citetalias{Meingast-II:2019} the stream's extent might potentially be much larger due to sensitivity limitations. The method used in this paper is not able to infer the stream membership of stars outside the constructed decision boundary. Finding externally located stream members would require, for example, a transition to unsupervised methods which are not limited by a fixed training set. 

Additionally, the constructed decision boundary depends heavily on the outermost points in the training sample as they are more likely to act as support vectors for the decision surface. As the density of points decreases towards these outer regions (in 5D) the decision boundary depends on random fluctuations of these border points present in the training set. Furthermore, we suspect the fraction of contaminants in stream member stars per unit volume to increase towards border regions. Thus, outliers in the border region have an increasing chance of being a support vector defining the shape of the decision surface. These effects, however, are somewhat mitigated by the choice of bagging multiple predictors, which helps to reduce unstable decision surfaces. 

While omitting the radial velocity component opens up the possibility to search for more stream members we  lose at the same time an additional discriminative dimension. By neglecting the radial velocity distribution of the input data, the implemented classification scheme impacts the contaminant fraction of our final source list. This results in a trade-off between increasing the overall recall at the expense of a reduced precision.

\section{Results and discussion}
\label{sec:results}
\begin{figure}[ht]
\centering\includegraphics[width=0.9\linewidth]{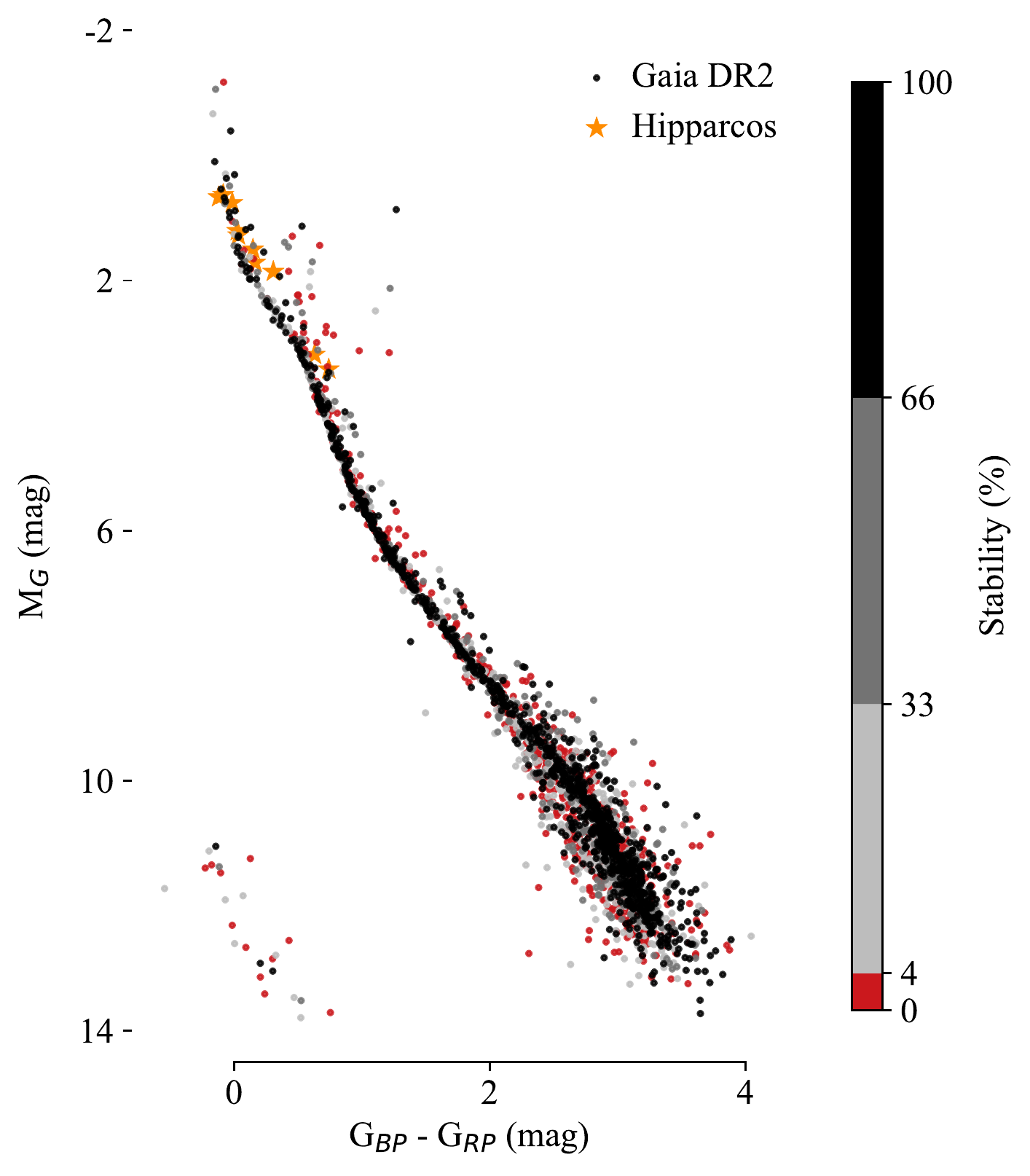}
\caption{Distribution of predicted sources in the color-absolute magnitude diagram. The shades of gray encodes the stability information of each source. The hue change in the color map at $4\%$ denotes the transition from robust stream members in gray tones to less reliable sources in red.
Additionally, we show $10$ new potential stream members, identified by applying the same classifier to the Hipparcos catalogue. 
}
\label{fig:hrd_sources_stability_cut}
\end{figure}

Using no pre-filter selection the classifier ensemble predicts a total of 4243 stream members. This source list does, however, not contain all members from the original training set. Approximately $10\%$ of the training data are not captured by the ensemble classifier. This reduction can be attributed to the model validation phase where we prioritize more conservative models in an attempt to prevent overfitting. To visualize our results we implement a series of quality selections described in App.~\ref{app:cut}, hereinafter referred to as filter Q1. For a direct comparison to the original training sample we implement the filter criteria as in~\citetalias{Meingast-II:2019} (excluding the criterion on radial velocities), hereinafter referred to as filter Q2. The quality filters Q1 and Q2 reduce the total number of classified member stars to $2567$ and $2913$, respectively. This selection contains, however, many sources which are predicted by only a marginal fraction of the $8515$ classifiers in the bagging ensemble. Each individual classifier is associated with an individual set of classified stream members. Thus, considering all 8515 classifiers, each source can be assigned a prediction frequency. We define this prediction frequency, hereinafter referred to as \emph{stability}, as the fraction of classifiers in the bagging ensemble that include a certain star in their prediction set. Fig.~\ref{fig:xyz_pm_scattermatrix} shows the five-dimensional distribution of the training sample (top row) and the stream members classified by our trained OCSVM (quality filter Q1), where the color indicates the stability of each source for our new classification. We observe that, on average, stability values tend to increase towards the central parts of the stream. 
Additionally, we find that when inspecting the new source set in the color-absolute magnitude diagram, see Fig.~\ref{fig:hrd_sources_stability_cut}, sources with lower stability numbers correlate with a larger scatter while sources with higher stability values are more compactly distributed around an idealized isochronal curve. Therefore, stability can be used as a measure to filter out potential contaminant sources. 

Since the training process includes a validation step even stars with low stability values can be regarded as potential stream members. Hence, stability constitutes not a probability estimate but rather a quality feature for which we aim to find a suitable criterion to clean our prediction sample. To determine the reliability of the predicted stellar sample we estimate the level of contamination at various stability filters.

We measure the contamination via the velocity dispersion in 3D, parametrized via $v_r$, $v_\phi$, and $v_z$. However, due to contributions of random contaminants the standard error of the prediction set is largely dominated by outliers, regardless of the stability filter criterion. Hence, we describe the variability of the velocity distribution with the median absolute deviation (MAD) which is a robust estimate of statistical dispersion. For reference, the training data distribution measures a MAD in the 3D velocities of \SI{2.1}{\km \per \s}.

\begin{figure}
\centering\includegraphics[width=0.95\linewidth]{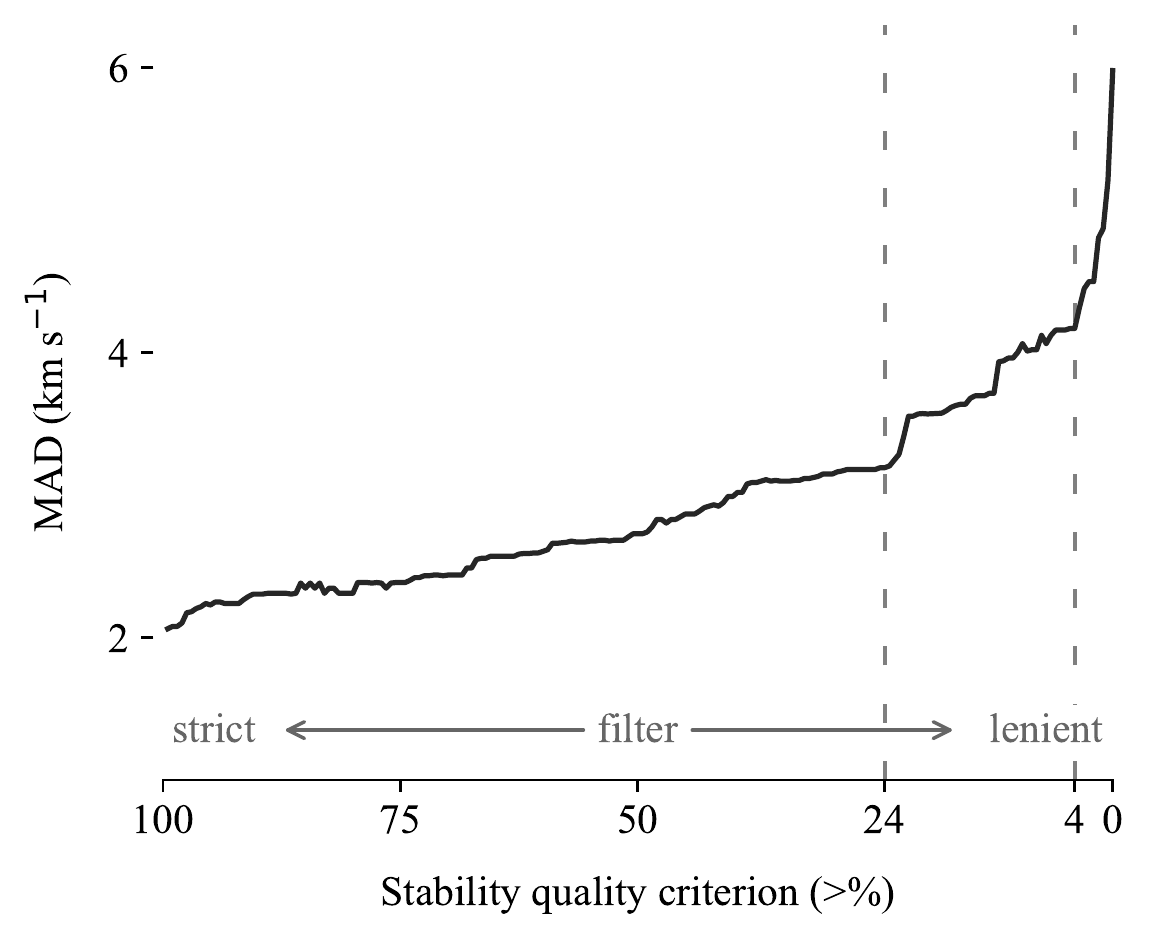}
\caption{Median absolute deviation of sources from the expected 3D velocity as a function of the stability quality filter. The x-axis is reversed displaying very strict filter criteria on the left most side and lenient filter criteria towards the right side. A trend is visible where the amount of scatter over the stability filter is split into two parts where each is characterized by a different slope. 
Suitable quality filters are realized by \texttt{stability > $4\%$} and, more conservatively, \texttt{stability > $24\%$}.
}
\label{fig:mad_stability}
\end{figure}

Fig.~\ref{fig:mad_stability} displays the influence of a variable stability filter criterion on the 3D velocity distribution. By moving in the plot from left to right we gradually add less "stable" sources to the predicted data set. We identify two distinct sections in this curve that are dominated by different slopes. Firstly, the section with stabilities from $100 \%$ decreasing to $4 \%$ is comprised of a roughly constant growing scatter around the expected 3D Cartesian velocity. Secondly, adding sources with a stability below $\sim 4\%$ results in a rapid growth of the MAD. This sudden increase 
is most likely caused by adding a significant number of contaminating field stars.
Here, we assume that these contaminating field stars are more likely associated with the outer borders of the stream in the 5-dimensional parameter space which is also where the trained classifier ensemble is less confident about the stream membership of stars. This decrease in stability values of predicted sources towards the outer regions of the stream is also well visible in Fig.~\ref{fig:xyz_pm_scattermatrix}.

In addition to the sudden increase at $4\%$, we identify another characteristic property of the MAD distribution in Fig.~\ref{fig:mad_stability}. Starting at about 40\% we observe an extended flat distribution up to 24\%. In this range the amount of scatter remains nearly constant. This filter criterion (\texttt{stability$\ge 24\%$}) yields a very stable subsample to the more lenient \texttt{stability>$4\%$} criterion.

\begin{figure*}[ht]
\centering\includegraphics[width=\linewidth]{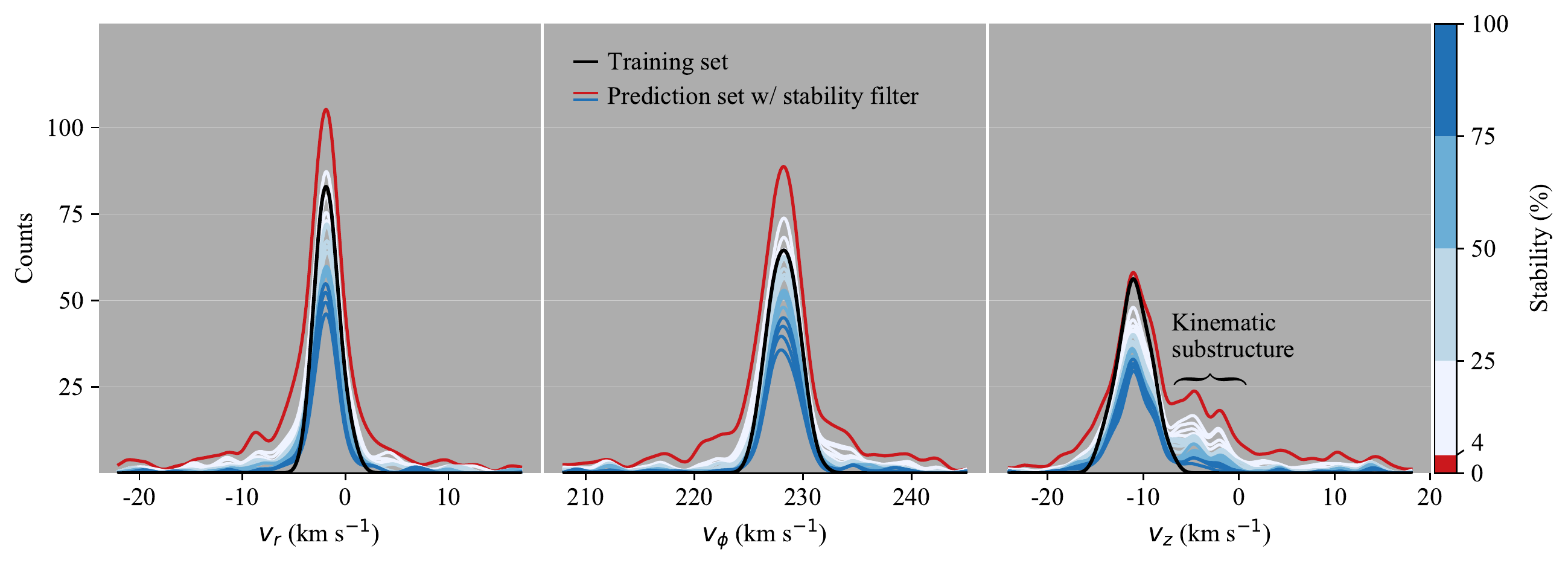}
\caption{Kernel density estimation of the marginal 3D velocity distributions for various stability filter criteria. The individual lines are color-coded by the filter criteria and range from red (\texttt{stability < 4\%}) to dark blue which represents the most strict filter criterion. The distributions are sampled at constant intervals in stability with a step size of 5\%. The hue change from red to shades of blue indicates the transition from the contamination dominated to the more robust filter regime. In addition, we note a kinematic substructure in the $z$-velocity distribution which is indistinguishable from other sources in all features except $v_z$.}
\label{fig:vel_stability_filter}
\end{figure*}

The filter behaviour can be observed in more detail in Fig.~\ref{fig:vel_stability_filter} where the successive cleaning of the prediction set is displayed in each individual velocity component. The solid lines in the figure represent a kernel density estimation of the marginal distributions for various color-coded stability filter criteria. Specifically, we sampled the distributions at constant intervals in stability with a step size of $5\%$. The hue change from red to shades of blue indicates the transition from a contamination-dominated to a more robust filter regime. In the marginal distributions the disproportionately large reduction in the amount of scatter around mean velocities by applying the \texttt{stability>$4\%$} filter criterion becomes apparent. For subsequent filter criteria the contamination outside the training sample distribution (black line) is reduced at a nearly constant rate, particularly in the $v_r$ and $v_\phi$ observables.
Moreover, we identify a kinematic substructure in the panel displaying $v_z$ velocities. Sources identified with this substructure have systematically larger vertical velocities by about \SI{5}{\km \per \s} compared to the bulk motion of the stream. These sources are only clearly separable in $v_z$ and do not show any obvious correlation in other velocities or can be segregated in spatial coordinates. We note here that this substructure accounts for the high MAD of the predicted sources and is removed only for very conservative stability filter criteria above $90\%$.

Following the above outlined characteristics in the velocity distributions, we therefore implement an additional criterion of \texttt{stability>$4\%$} or \texttt{stability>$24\%$} for a more conservative approach. Depending on the quality filter selection, the stability >$4\%$ filter criterion reduces the number of predicted stream members to $1869$ or $2110$ for Q1 and Q2, respectively.

\begin{figure}[ht]
\centering\includegraphics[width=0.95\linewidth]{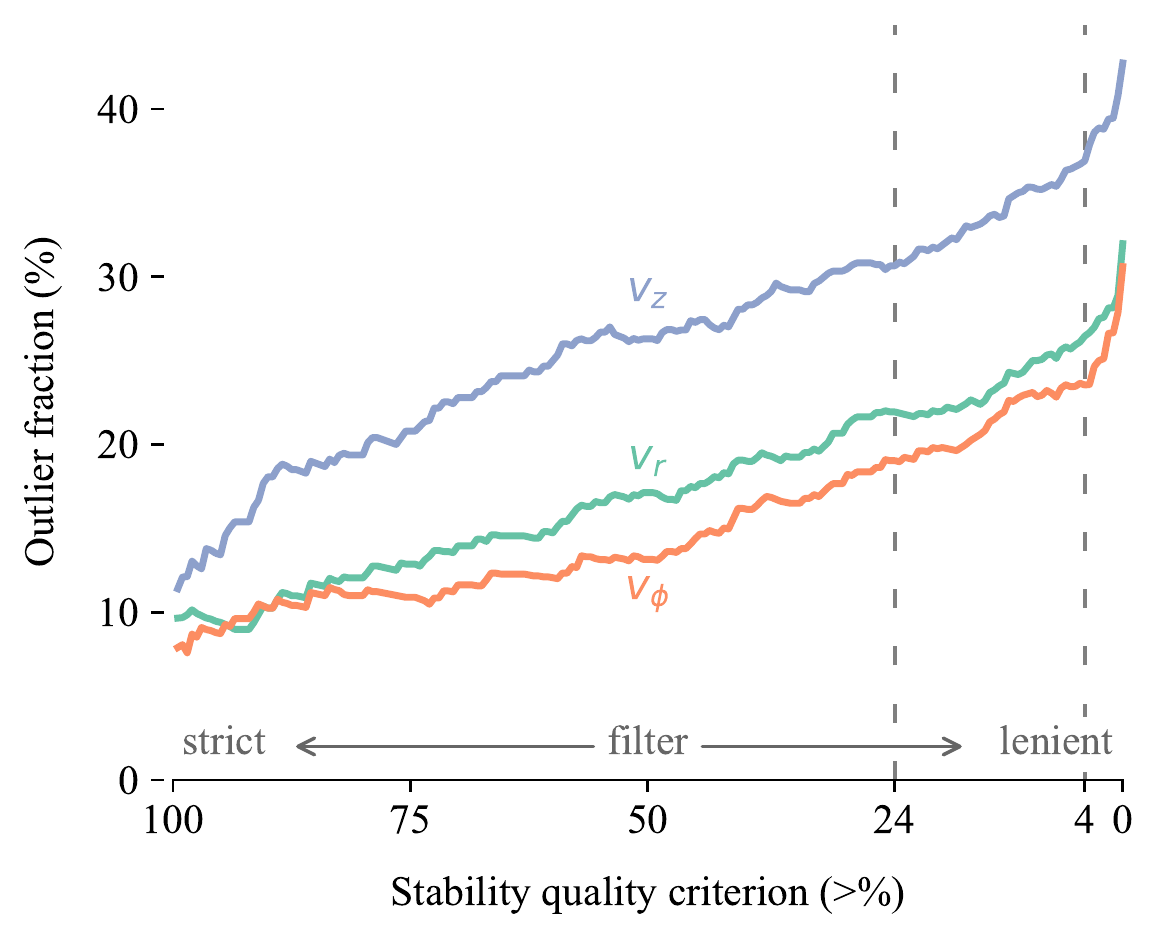}
\caption{Outlier fraction in the individual velocity components for a variable stability filter criterion. Due to a newly identified kinematic substructure in $v_z$ we estimate the contamination only in the radial and azimuthal velocity components (see Sec.~\ref{sec:results}). Based on this premise, the contamination is estimated to be roughly $25\%$ and $20\%$ for the stability criteria $>4\%$ and $>24\%$, respectively.}
\label{fig:outlierfrac_stability}
\end{figure}

In order to quantify the contamination fraction in our source catalog we consider the fraction of outliers in the marginal 3D velocity distributions. To do so, we define for each velocity component a region of inliers as the $3 \sigma$ around the training sample mean. This definition constitutes a very conservative estimate as the velocity distribution of the training data is by design very narrow. Furthermore, the kinematic substructure in the $v_z$ component naturally leads to very large contamination fractions. 
For this reason, we only consider the radial and azimuthal velocity components when estimating the contamination for various stability filter criteria. Fig.~\ref{fig:outlierfrac_stability} shows the outlier fraction within each velocity component. Based on our assumptions we obtain a contamination estimate of roughly $25\%$ and $20\%$ for the stability criteria $>4\%$ and $>24\%$, respectively. However, we note again that this is a very conservative estimate which assumes an intrinsic velocity dispersion of only around \SI{1}{\km \per \s}. By increasing the estimated velocity dispersion to \SI{2}{\km \per \s} the contamination drops to roughly $10 - 15\%$ which we suspect to be a more realistic estimate.

Since the ensemble classifier is trained on positional and proper motion data we can apply it to any survey which provides these measurements. In an effort to increase the source list especially towards brighter stars we apply our ensemble classifier to the Hipparcos~\citep{hipparcos_new:2007} source catalogue, see App.~\ref{app:hipparcos} for more details. In total, we find 21 new potential stream members in the Hipparcos catalogue, 10 of whom we consider to be robust. We have added the 10 predicted Hipparcos sources to the HRD plot in Fig.~\ref{fig:hrd_sources_stability_cut}. Among the prediction set we find $\alpha$ Aquarii, the brightest star in the Aquarius constellation. Using the radial velocity information from~\citet{Soubiran_2008} we find a galacto-centric velocity of $\Vec{v} = (-3.15, 229.19, -8.73)$ km s$^{-1}$, which is well within the $3 \sigma$ region of the training set. However, a comparison of parallax measurements between Gaia and Hipparcos reveals a large systematic discrepancy of a factor of approximately two which makes $\alpha$ Aquarii a low-fidelity stream member.

\begin{figure*}[ht]
\centering\includegraphics[width=0.95\textwidth]{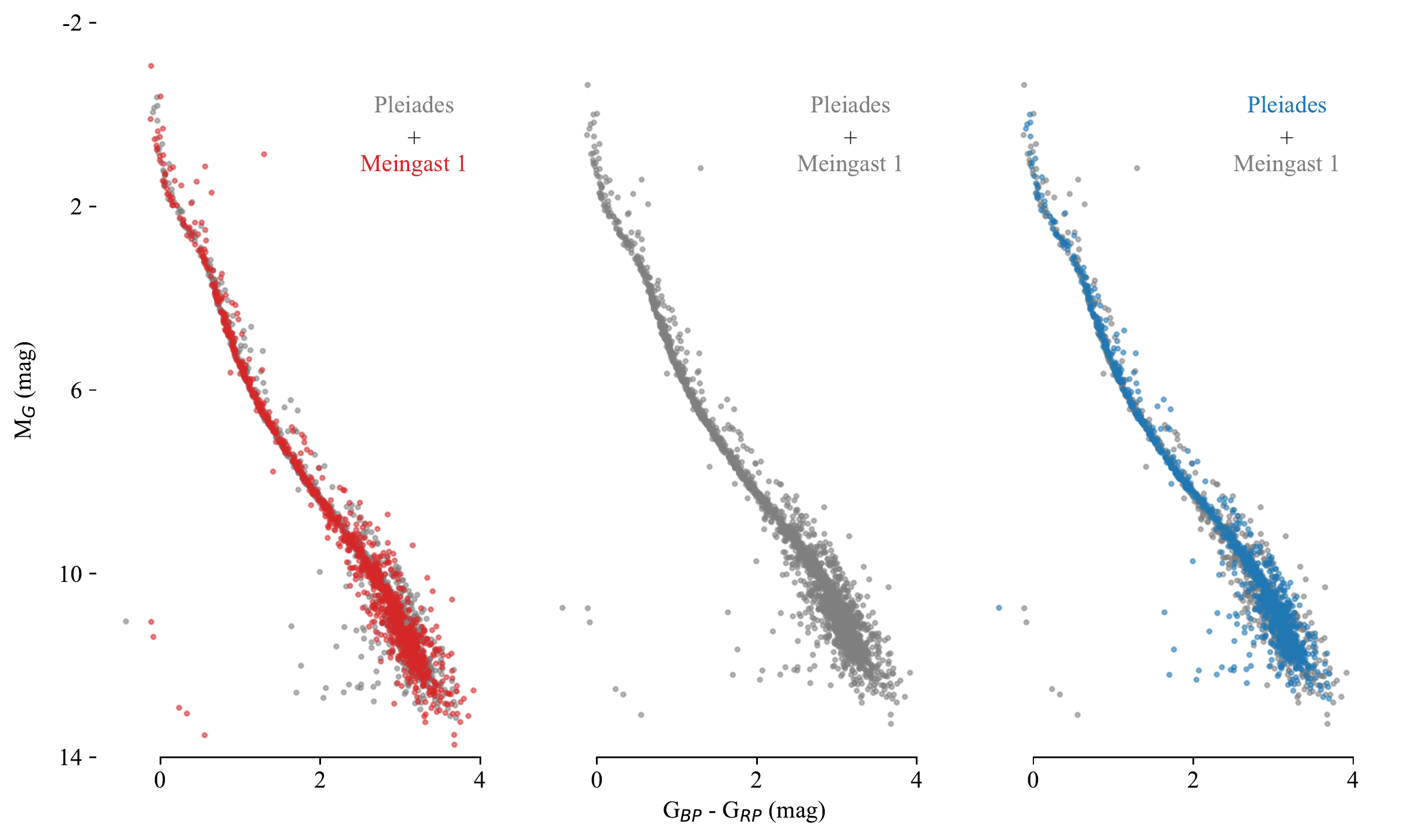}
\caption{Comparison between the predicted stream members and the Pleiades member selection. The three panels show the same two data sets plotted on top of each other and highlighted by different colors. In the left plot the Pleiades are highlighted in blue while the predicted stream members are kept in grey. The center plot displays both stellar associations in grey. The right plot displays the predicted stellar stream in red on top of the Pleiades in grey. We chose to the stability cut to match the number of sources in the Pleiades sample in order to generate a fair comparison. The CMD-distributions of the Pleiades and the predicted stream matches almost perfectly.}
\label{fig:cmd_comparison}
\end{figure*}

\subsection{Age and mass determination}

Using gyrochronology, \cite{Curtis:2019} concluded that the stream has an age comparable to the Pleiades. This contrasted with the isochronal age derived in \cite{Meingast-II:2019}, which was hinging on a single star, 42 Ceti, a sub-giant.
With the new and larger member list we can now attempt to make a more precise estimate regarding the stream's age. 

We compare the stream to a selection of the Pleiades members~\citep{PleiadesSources:2018}. By introducing a slight color offset of (G$_{BP}$ - G$_{RP}$ + 0.03) to the stream we find that the source distributions in the HRD of the Meingast 1 stream and the Pleiades match almost perfectly, as seen in Fig.~\ref{fig:cmd_comparison}, implying a similar age between the two stellar systems. The small color shift could imply either the presence of dust extinction towards the Pleiades or a lower metallicity of the stream, or both. The Pleiades are known to be affected by small amounts of extinction. Additionally, we find a slight metallicity difference between the stream and the Pleiades measured by LAMOST~\cite{LAMOST_paper} which is illustrated Fig.~\ref{fig:lamost}. The plot shows a discrepancy between the mean metallicity fraction of the two stellar populations where sources in Meingast 1 appear to be slightly more metal-poor than the ones in the Pleiades which could help to explain the reddening in color space. 

The three panels in Fig.~\ref{fig:cmd_comparison} show the source distributions in the HRD of both, the Meingast 1 stream and the Pleiades, plotted on top of each other and highlighted by different colors. In the left plot sources in the Meingast 1 stream are highlighted in red while the Pleiades members selection are kept in grey. The center plot displays both stellar populations are shown in grey. The right plot displays the Pleiades in blue on top of Meingast 1 in grey. In order to do a fair comparison we define the stability filter in such a way that the number of sources of the stream is equal to that of the Pleiades. This results in the following filter criterion: $\texttt{stability>45.9}$. The particular similarity of the two distributions suggests an approximately identical age. The Gaia collaboration~\citep{ObservationalHRD:2018} estimates the age and metallicity fraction of the Pleiades to be $110$~My and Z=$0.017$, respectively. Therefore, our age estimate is, within the expected error range, consistent with~\citet{Curtis:2019}.

\begin{figure}[ht]
\centering\includegraphics[width=0.95\linewidth]{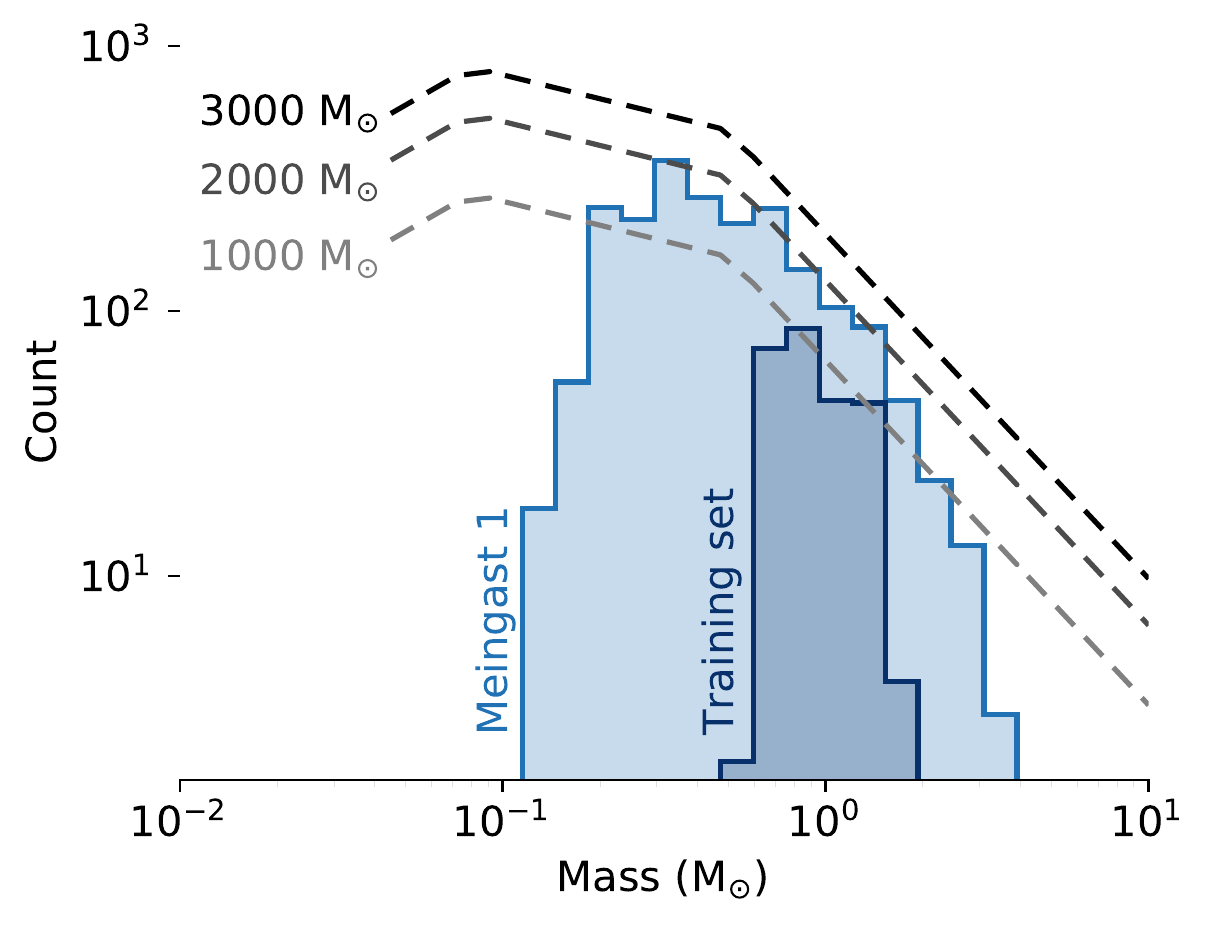}
\caption{Mass function for the Meingast 1 stream sources (light blue) and the training examples (dark blue). The dotted lines indicate model IMFs within a cluster massrange of $1000-3000$ M$_{\odot}$.}
\label{fig:mass}
\end{figure}

We estimate the total mass of the selected sources in accordance with~\citetalias{Meingast-II:2019} by using PARSEC isochrones. Using an age estimate of $110$ My and a metallicity fraction of Z=$0.016$ results in a mass distribution shown in Fig.~\ref{fig:mass}. The plot depicts the mass distribution of the training samples (dark blue) versus the predicted samples (light blue). The dotted gray lines indicate IMFs~\citep{Kroupa:2001} for clusters masses of $1000$ M$_{\odot}$, $2000$ M$_{\odot}$, and $3000$ M$_{\odot}$. A comparison to the model IMFs suggests an approximate mass of $2000$ M$_{\odot}$, as suggested in~\citetalias{Meingast-II:2019}. To our knowledge, this makes the Meingast 1 stream the most massive stellar stream in the solar neighborhood. 

\section{Summary and conclusion}
We revisit the stream discovered in~\citet{Meingast-II:2019} to search for new members using Gaia DR2 data and a machine learning approach. Using the original source selection as training data we deploy a bagging classifier of one-class Support Vector Machines to the full Gaia DR2 data searching for new stream members in position and tangential velocity space. The ensemble classifier is created in a hyper-parameter search combined with a model selection that rejects models which do not meet a set of preconditions. The resulting set of classifiers creates a variable prediction frequency for possible stream member stars which we use as a criterion to select high fidelity sources. Subsequently, we validate the newly found sources in the HR diagram and the Cartesian velocity distribution.

In total, we find about 2000 stream member stars with high-fidelity increasing the source population approximately tenfold. As the newly predicted stream members are no longer limited by radial velocity measurements the new selection substantially extends the main sequence unveiling the stream’s population across the entire stellar mass spectrum, from B-stars to M-stars, including white dwarfs. In a comparison in the color-absolute magnitude diagram we find that, apart from being slightly more metal-poor, the stream is indistinguishable from that of the Pleiades cluster, suggesting a similar age. In the mass range at which we are mostly complete, $\sim 0.2$ < M$_\odot$ < $\sim 4$ M$_\odot$, we identify a normal IMF. This comparison allows us to estimate the total mass of the stream to approximately 2000 M$_\odot$, making it by far the most massive stream we know. Additionally, we find several white dwarfs as members of the stream.


\begin{acknowledgements}
This work has made use of data from the European Space Agency (ESA) mission Gaia (\href{https://www.cosmos.esa.int/gaia}{https://www.cosmos.esa.int/gaia}), processed by the Gaia Data Processing and Analysis Consortium (DPAC, \href{https://www.cosmos.esa.int/web/gaia/dpac/consortium}{https://www.cosmos.esa.int/web/gaia/dpac/consortium}). Funding for the DPAC has been provided by national
institutions, in particular the institutions participating in the Gaia Multilateral
Agreement. This research made use of Astropy, a community-developed core
Python package for Astronomy~\citep{astropy:2018}. We also acknowledge the various Python packages that were used in the data analysis of this work, including NumPy~\citep{van_der_Walt:2011}, SciPy~\citep{Scipy}, scikit-learn~\citep{scikit-learn}, Pandas~\citep{mckinney-proc-scipy-2010}, and Matplotlib~\citep{matplotlib:2007}. This research has made use of the SIMBAD database operated at
CDS, Strasbourg, France~\citep{simbad}
\end{acknowledgements}

%
%
\newpage
\bibliographystyle{apalike}
\bibliography{sources}

\begin{appendix}
\section{Training process}
\label{app:training}
The training of each individual predictor in the full model ensemble is summarized in the following two steps.

First, we select a random pair of hyper-parameters ($\gamma_{i}$, $\nu_{i}$, $(c_x/c_v)_i$) and train a model with 10 fold Cross Validation (CV). Due to a contamination of field stars of a few percent~\citetalias{Meingast-II:2019} we encourage stricter and more compact descriptions of the stream (in 5D) ignoring potential outliers in the training sample. In a first selection step we filter models with a low average accuracy across the holdout sets of $<0.5$ or a standard deviation of above $0.15$. The standard deviation filter helps to obtain fairly conclusive predictors for different sub-samples on a fixed set of hyper-parameters.

Second, models which pass the CV step are deployed on the full data set (see Sec.~\ref{sec:data}). In an effort to minimize contamination of nearby\footnote{Nearby is referring to sources being in the vicinity to the stellar stream in the 5D feature space.} field stars and thus boosting robustness of the prediction we train the model on $10$ bootstrap samples, with a sample size of $80\%$ of the training data size. The union of all $10$ predictions is then considered the final model. Before we add the newly trained model (with the hyper-parameter set ($\gamma_{i}$, $\nu_{i}$, $(c_x/c_v)_i$) into the final bagging classifier we validate its performance against our prior beliefs about the approximate model structure described in Sec.~\ref{sec:hp_tuning}

\section{Validation process}
\label{app:validation}
After training a classifier we validate its ability to capture important physical aspects about the estimated size and shape of the stellar stream. We require the classifier to capture at least the following criteria:
\begin{enumerate}
\item The number of predicted stream members $N_s$ must not exceed a physically sensible range which is limited to $N_s \in [500, 5000]$.
\item The extent of the predicted stream members in position and proper motion space must be similar to the original ones.
\item TThe cylindrical velocity distribution of the stream members must not deviate too much from the training sample distribution.
\end{enumerate}
The similarity condition (2.) is achieved by requiring the dispersion of the predicted to the original stream members in position and proper motion space to be approximately equal. We approximate the extent, or dispersion of the stream in both spaces by a single number, namely the mean distance $\overline{d}$ of its member stars to the centroid of the full stream. For a point in position space $\Vec{r} = (x, y, z)$ and its corresponding centroid $\Vec{r}_{c}$, $\overline{d}_{\Vec{r}}$ is
\begin{equation}
    \overline{d}_{\Vec{r}} = \frac{1}{N} \sum_{i}^{N} ||\hspace*{1pt}\Vec{r}_{i} - \Vec{r}_{c}||,
\end{equation}
where $N$ is the number of stars belonging to the cluster. Respectively, in proper motion space with a point $\Vec{v} = (\mu_{\alpha}, \mu_{\delta})$ and centroid $\Vec{v}_{c}$, $\overline{d}_{\Vec{v}}$ is:
\begin{equation}
    \overline{d}_{\Vec{v}} = \frac{1}{N} \sum_{i}^{N} ||\hspace*{1pt}\Vec{v}_{i} - \Vec{v}_{c}||.
\end{equation}
We use these two structure parameters $\overline{d}_{\Vec{r}}$ and $\overline{d}_{\Vec{v}}$ to determine the extent of the stream in position and proper motion space, respectively. Our aim is to find models whose predicted points retain a similar dispersion as the original ones. To avoid overfitting we compare the dispersion of the prediction set to the training set which acts as an upper limit:
\begin{equation}
  \overline{d}_{\Vec{r}/\Vec{v}}^{\text{orig}} > \overline{d}_{\Vec{r}/\Vec{v}}^{\text{pred}}
\end{equation}

Lastly, we control the centroid position of the predicted stream members to avoid systematic shifts. The predicted and original stream centroid must be reasonably close to each other with respect to the average dispersion of training points.
  \begin{equation}
  || \hspace*{1pt}\vec{r}_{c}^{\text{org}} - \vec{r}_{c}^{\text{pred}} \hspace*{0.5pt}|| < \overline{d}_{\Vec{r}}^{\text{org}} \times 0.1
  \end{equation}
  
  \begin{equation}
  || \hspace*{1pt}\vec{v}_{c}^{\text{org}} - \vec{v}_{c}^{\text{pred}} \hspace*{0.5pt}|| < \overline{d}_{\Vec{v}}^{\text{org}} \times 0.1
  \end{equation}
 
 The third condition is implemented by examining the contamination of predicted samples compared to the training sample. To get a rough estimate of the contamination we compare the galacto-centric velocity distribution, \emph{i.e.} $\Vec{v}=(v_r, v_\phi, v_z)$, of the predicted sources to the training sample. Instead of comparing the velocity dispersion of both samples we characterise the level of contamination by considering the fraction of outlier sources. This way, we try to mitigate the influence of large outliers which increase the dispersion drastically for such a low number of sources. In order to characterize outlier sources we consider the training examples. Assuming that almost all sources lie within the $\pm 3\sigma$ range around the mean we consider the ratio of sources lying outside of the $3 \sigma$ range compared to the total amount of sources. A classifier is rejected if on average, across the individual velocity components, more than $25\%$ of sources are considered outliers. The aim of this criterion is to remove models which extend into a region of feature space where the radial velocity distribution does not match our assumption of a co-moving structure.

\section{Parameter tuning results}
\label{app:hp}
\begin{figure*}[ht]
\centering\includegraphics[width=0.95\linewidth]{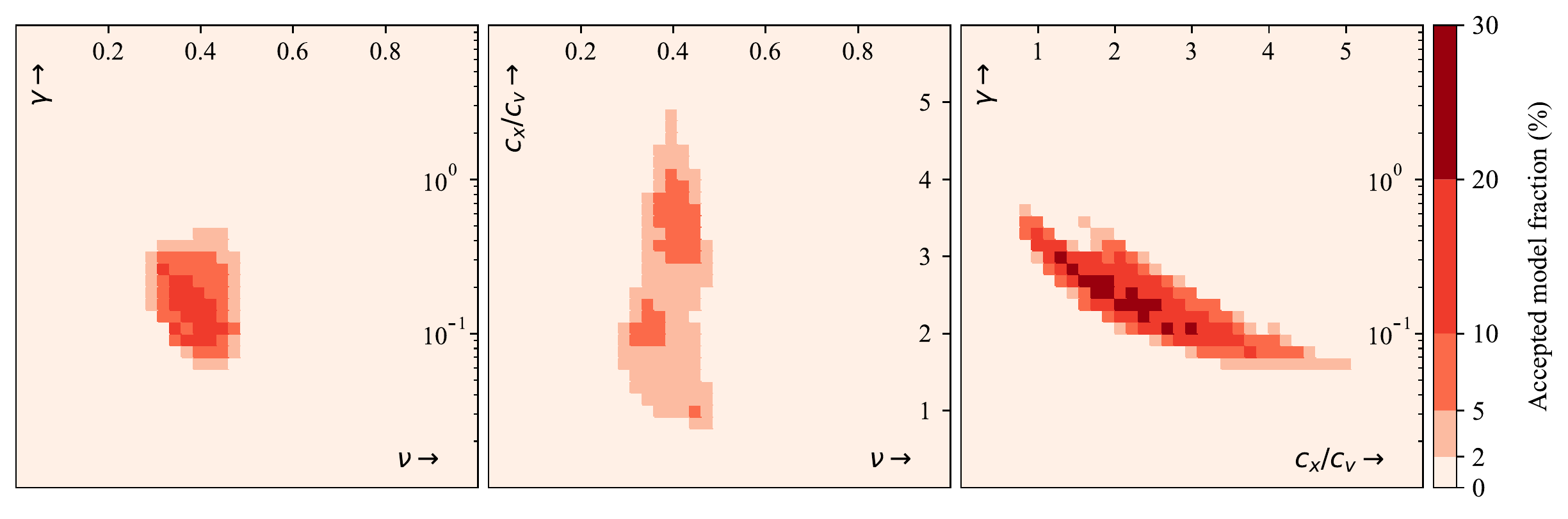}
\caption{Hyper-parameter search in the parameters $\nu$, $\gamma$, and $c_x/c_v$ yielding the one-class support vector machine bagging predictors. The color information illustrates the accepted model faction within a certain hyper-parameter bin range. A classifiers is accepted if it passes the quality criteria presented in Sec.~\ref{sec:hp_tuning}. The model ensemble consists of $8515$ individual predictors.}
\label{fig:hp_selected}
\end{figure*}

The hyper-parameter search in combination with a classifier selection and validation step, see Sec.~\ref{sec:hp_tuning} yields a set of approved parameter triples ($\nu_i$, $\gamma_i$, $(c_x/c_v)_i$) which make up the final OCSVM bagging predictor. The distribution of accepted triples is displayed in Fig.~\ref{fig:hp_selected}. The color information illustrates the accepted model faction within a certain hyper-parameter bin range. A model is accepted if it passes the quality criteria presented in Sec.~\ref{sec:hp_tuning}. The model ensemble consists of $8515$ individual predictors. 

\section{Quality criteria}
\label{app:cut}
In general the source identification method we present in this paper is independent of any quality criteria. However, in order to show the distribution of stars in the color magnitude diagram we apply the following error criteria on data quality. Following the description in~\cite{Lindegren:2018} the five-parameter solution depends on the number of visibility periods used for a certain source. A visibility period is defined as a group of observations separated from other groups by a gap of at least four days. Since a five-parameter solution is accepted only for \texttt{visibility\_periods\_used}>6 we implement said criterion. 

A recommended astrometric quality parameter is the re-normalised unit weight error (RUWE) described by~\cite{Lindegren_RUWE:2018}. It is based on a re-calibration of the unit weight  error described in~\cite{Lindegren:2018}. We follow the advice in the technical note~\citep{Lindegren_RUWE:2018} and use the criterion \texttt{RUWE}<1.4 to select astrometrically reliable sources. Furthermore, we implement additional astrometric quality measures, \texttt{astrometric\_sigma\_5D\_max}<0.5 and $\varpi / \sigma_{\varpi}$>10, which reduce the number erroneous measurements.

Finally, we adopt the following photometric quality criteria, \texttt{phot\_bp\_mean\_flux\_over\_error}>10 and \texttt{phot\_rp\_mean\_flux\_over\_error}>10.

\section{Metal content}
\label{app:metal}
\begin{figure}[ht]
\centering\includegraphics[width=0.95\linewidth]{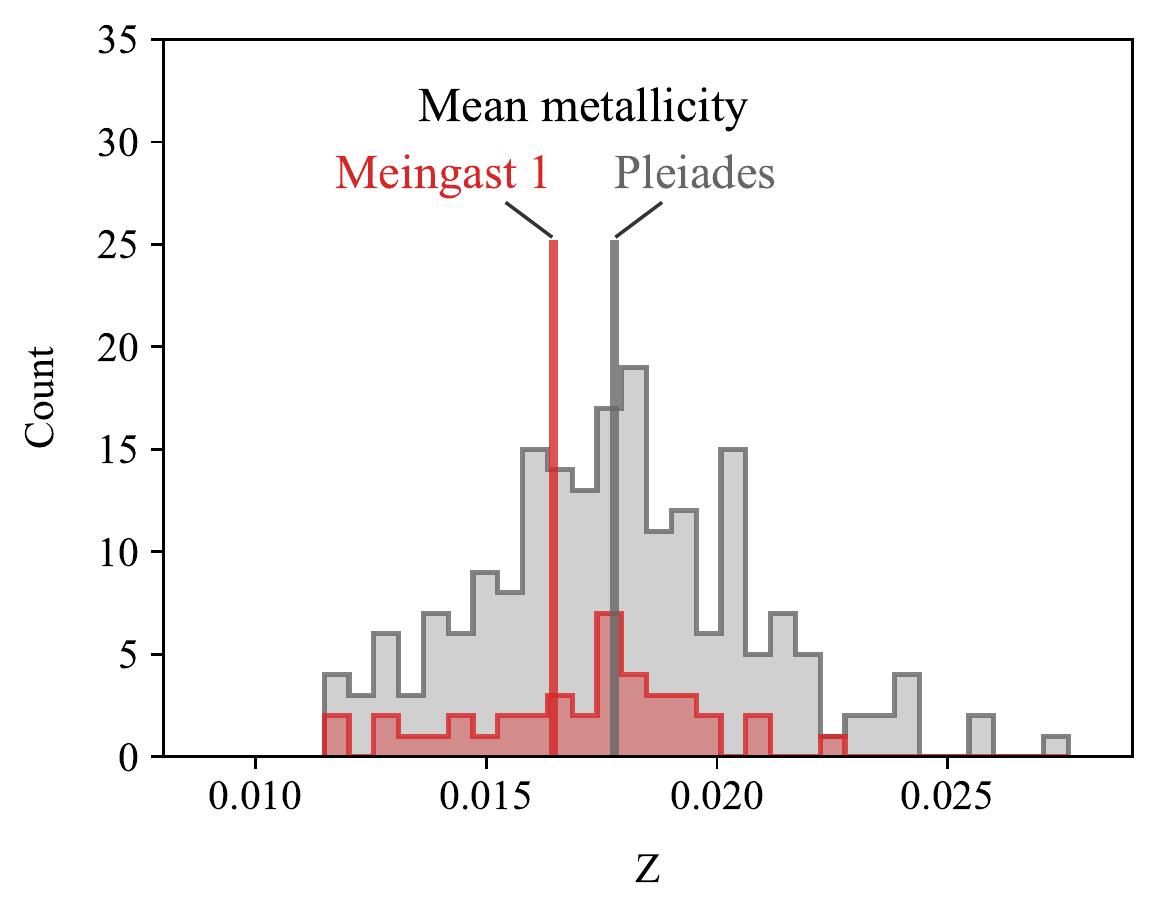}
\caption{Comparison of the metallicity fraction of Pleiades and Meingast 1 memeber stars. The vertical lines indicate the mean metal content of both populations. We find that the members of the Meingast 1 associtation are slightly more metal poor than the Pleiades.}
\label{fig:lamost}
\end{figure}

Fig.~\ref{fig:lamost} shows a comparison of the metallicity fraction Z between a Pleiades member selection~\citep{PleiadesSources:2018} and the stream members. A cross-match of the Pleiades and stream source selections to the LAMOST DR5~\cite{LAMOST_paper} catalogue results in $383$, and $83$ matches, respectively. The conversion from chemical abundance ratios [Fe/H] to the metallicity fraction Z has been made in accordance to the PARSEC~\citep{PARSEC:2012} solar value of Z $=0.015$. Subsequently, we filter out the most untrustworthy sources by requiring that the error of the measured chemical abundance ratios [Fe/H] is below $0.05$ and [Fe/H] $>-1$. Additionally, we only select sources above an effective temperature of $5000$ K. These criteria yield $197$ and $44$ matched sources for the Pleiades and the Meingast 1 stream, respectively. The metal content distributions of the Pleiades and stream members show a large scatter but the positions of their respective mean indicate that the Meingast 1 stream members appear to be slightly more metal poor compared to the Pleiades member stars. 

\section{Hipparcos source selection}
\label{app:hipparcos}
Compared to the training samples from the Gaia DR2 catalogue the Hipparcos sources have larger associated standard errors of measured quantities. Considering the higher uncertainty in the Hipparcos catalogue variables we adopt a more conservative stability filter criterion of \texttt{stability}$>50\%$. Despite a rather high stability cut a large standard error increases the chance of contaminant stars falling into the selection. Therefore, we adopt a second quality filter where we sample each data point from marginal normal distributions centered on the provided mean value with a standard deviation of the provided standard error of each observable. We then draw $100$ samples per source from these marginal distributions and count how often these re-sampled sources are again predicted to be a stream member with \texttt{stability}$>50\%$. Eventually, this quality criterion yields $11$ additional sources with a re-sampling fraction of over $50\%$.

\section{Table content}
The content of the published source catalogue is summarized in the Table~\ref{table:source_list}.
\begin{table}[h!]
\centering
\begin{tabular}{c c} 
 \hline \hline \\[-1.4ex]
 Column name & Description \\ [0.5ex] 
 \hline \\[-1.7ex]
 source\_id & Gaia DR2 source identification number \\ 
 ra & R.A. (deg) \\
 dec & Declination (deg) \\
 X & $x$-Position (pc) \\
 Y & $y$-Position (pc) \\ 
 Z & $z$-Position (pc) \\
 pmra & $\mu_\alpha$ (mas yr$^{-1}$) \\
 pmdec & $\mu_\delta$ (mas yr$^{-1}$) \\
 stability & Stability percentage (\%) \\
 q1 &  Filter criterion Q1 (bool); see App.~\ref{app:cut} \\
 q2 &  Filter criterion Q2 (bool); see~\citetalias{Meingast-II:2019} \\[1ex]
 \hline\\[-1ex]
\end{tabular}
\caption{Contents of the source catalogue which are available online via CDS. The positional data XYZ are measured in Galactic Cartesian coordinates centered on the sun.}
\label{table:source_list}
\end{table}

\end{appendix}

\end{document}